\documentclass[submission,copyright,creativecommons]{eptcs}
\providecommand{\event}{Third Workshop on Agents and Robots for reliable \\ Engineered Autonomy} 

\usepackage{iftex}
\usepackage{graphicx}
\usepackage{subcaption}
\usepackage{algorithm}
\usepackage{algpseudocode}
\usepackage{multicol}
\usepackage[compact]{titlesec}
\usepackage{todonotes}
\usepackage{soul}
\usepackage{svg}
\usepackage{multirow}
\usepackage{amsmath}
\usepackage{adjustbox}
\usepackage{colortbl}

\ifpdf
  \usepackage{underscore}         
  \usepackage[T1]{fontenc}        
\else
  \usepackage{breakurl}           
\fi

\newcommand{\mytitle}{Evaluating Heuristic Search Algorithms in Pathfinding: A Comprehensive Study on Performance Metrics and Domain Parameters}

\title{\mytitle{}\thanks{M. Robol and M. Roveri are partially supported by the project MUR PRIN 2020 - RIPER - Resilient AI-Based Self-Programming and Strategic Reasoning - CUP  E63C22000400001. M. Roveri and P. Giorgini are partially supported by the PNRR project FAIR - Future AI Research (PE00000013),  under the NRRP MUR program funded by NextGenerationEU.}}

\newcommand{\myspace}{\hspace*{1cm}}
\author{Aya Kherrour \myspace Marco Robol \myspace  Marco Roveri \myspace  Paolo Giorgini
\institute{Department of Information Engineering and Computer Science\\University of Trento -- Trento, Italy}
\email{name.surname@unitn.it}
}
\def\titlerunning{Evaluating Heuristic Search Algorithms in Pathfinding}
\def\authorrunning{A. Kherrour, M. Robol, M. Roveri, and P. Giorgini}

\begin{document}
\maketitle

\begin{abstract}
    The paper presents a comprehensive performance evaluation of some heuristic search algorithms in the context of autonomous systems and robotics.
    The objective of the study is to evaluate and compare the performance of different search algorithms in different problem settings on the pathfinding domain.
    Experiments give us insight into the behavior of the evaluated heuristic search algorithms, over the variation of different parameters: domain size, obstacle density, and distance between the start and the goal states.
    Results are then used to design a selection algorithm that, on the basis of problem characteristics, suggests the best search algorithm to use.
\end{abstract}


\section{Introduction}
    Autonomous agents and robotics have increasingly been used in various domains, such as, industrial applications \cite{muller2021industrial, perez2021introducing, javaid2021substantial}, surveillance \cite{pechoucek2008defence}, and exploration \cite{brat2006verification}. These systems are designed to autonomously make decisions and execute actions based on their dynamic and unpredictable environments. Under such conditions, systems are required to be as reactive as possible to changes in the environment. Therefore, ensuring good performance is a significant challenge. In the case of planning, selecting the most effective search algorithm becomes crucial to enhance the overall performance of the system. 
    Real-time heuristic search (RTS) (e.g.,\cite{korf1990real, koenig2001agent}) is a state-of-the-art approach for planning while executing, that helps minimize agent reaction time. These algorithms enable agents to make decisions by interleaving planning with execution while considering the evolving environment, which is an essential property in applications such as robotics, and video game agents. Despite the numerous methods proposed in this field \cite{korf1990real, koenig2009comparing, koenig2006real, bond2010real, DBLP:conf/ijcai/BjornssonBS09}, a comprehensive understanding of these algorithms remains elusive. Existing studies \cite{arica2017empirical, koenig2009comparing, realcomparison} have primarily focused on testing the performance of the algorithm based on a single parameter (e.g., look-ahead, sensor range). However, the influence of the problem domain characteristics on algorithm performance remains an understudied aspect.
    
    Our objective is to investigate the characteristics of the problem that may impact algorithm performances. To achieve this, we first review existing state of art of search algorithms and their applications. Then we design our experiments to evaluate some of the algorithm performances, also defining relevant metrics to use in the evaluation. Later on, experiment results are analyzed to provide a comprehensive understanding of how problem characteristics influence the performance of the different search algorithms. Finally, from the insight gained from our study, we introduce a selection algorithm that helps us to select the appropriate search algorithm.
    
    This paper is structured as follows. We begin by reviewing state-of-the-art search algorithms and previous performance evaluation studies. Next, we describe the approach we adopted to evaluate the search algorithms that are subject to our study, define the problem domain and its characteristics, and the performance metrics. In section four, we present our experimental results. Section five introduces our proposed selection algorithm with an execution example. Finally, we derive our conclusion, the limitations of our study, and possible directions for future work.    

\section{State of the art and related work}
    Path planning is a relevant problem in autonomous agents, such as, self-driving cars, robots, unmanned aerial vehicles (UAVs), and unmanned ground vehicles (UGVs), in which the host agent deliberates its path by moving from one position to another while avoiding obstacles and respecting some constraints \cite{sanchez2021path}. One of the path planning approaches that have been proposed to control the movement of these agent-based systems is search-based algorithms, with Dijkstra and its extension A*  \cite{DBLP:books/daglib/0023376, yershov2011simplicial}, being the most popular and effective ones. Besides these, other search algorithms have been proposed in the literature, specifically to reduce reaction time, broadly classified as real-time or incremental search algorithms.
    
    Real-time search algorithms must find a solution in a limited time, while incremental search instead uses the previously obtained searches to speed up the search. Amongst the first class, Real-time A* (RTA*) and Learning real-time A* (LRTA*) \cite{korf1990real} were some of the first algorithms to apply real-time heuristic search in path planning problems for moving agents. Both algorithms use heuristics to guide the search toward the goal, with RTA* storing the second-best f-values of the previous state as the best alternative to choose when backtracking from the current state. However, this may mislead the agent. Thereby, LRTA* overcomes this by storing the first best value rather than the second and learning from comparing the heuristic values of the adjacent states, thus preventing the algorithm from misleading the agent. 
    Another optimized version of LRTA* is real-time adaptive A* (RTAA*) \cite{koenig2006real}. It first determines its local search space and then speeds up the search by updating the heuristics values of states. It was developed for stationary target search problems and it follows trajectories of smaller costs. Anytime repairing A* (ARA*) \cite{likhachev2003ara} is a variation of A* that has been designed to find suboptimal solutions fast and then improve them over time, which makes it a good algorithm for problems where finding an optimal solution is not mandatory or too expensive \cite{likhachev2003ara}. However, finding suboptimal solutions does not make it find the optimal solution \cite{likhachev2003ara}.

    Moving to incremental search algorithms, D* (Dynamic A*) is an incremental search algorithm used in real-time planning in robotics \cite{stentz1995focussed}. It is designed to react quickly to changes in the environment, by updating the nodes affected in the search tree rather than recomputing a new plan from scratch. D* has a main drawback that it requires a lot of memory to perform the search. 
    LPA* (Lifelong Planning A*), an incremental version of A*. This algorithm is used in path planning or robot navigation in unknown terrain. It behaves just like A* in the first run, and then for the rest of subsequent searches it reuses the previous search thus reducing the number of examined nodes, which makes it fast. LPA* differs from D* in its search direction where it finds a path from the initial state to the initial goal state, therefore it does not fit in applications where the starting point may change over time. Another variant of D*, D* Lite, was developed based on LPA*, and is used for goal-directed path planning in unknown environments using the same idea as D*, however, it is a simple version of D* and produces effective results as the one delivered by D* as proven in \cite{koenig2002d}.
    While these heuristic search algorithms are numerous and diverse, it is important to emphasize that there are even more algorithms in the literature with new ones being developed, or existing ones being improved such as \cite{DBLP:conf/ijcai/BjornssonBS09, koenig2009comparing, hernandez2011real}. This variety of algorithms in the field emphasizes the complexity of the problem. Consequently, this variety of proposed algorithms presents a challenge; the wide number of these algorithms results in disparate performances in tasks such as path planning, making it difficult to select the optimal algorithms for a given task.
    
    Given the multitude of algorithms, studies have been done to compare the performance of RTS algorithms in path planning tasks such as \cite{arica2017empirical} where authors have compared the performances of path planning algorithms using agents equipped with a sensor in both stationary and moving target settings to select the most appropriate algorithm. The obtained results were compared using various sensor ranges and angles, and the authors concluded that sensor range is an important parameter in selecting the algorithm, unlike the sensor angle. Therefore the most appropriate algorithm can be selected based on the sensor range and its priorities. However, the authors did not consider varying the environmental characteristics, which may influence the performance of the algorithms.
    
    Another study has been presented in \cite{koenig2009comparing}, in which the authors also compared real-time and incremental heuristic search algorithms using an autonomous agent in a navigation task to know which class of heuristic search approach is better to be used depending on how informed the h-values are, and the search objective such as minimizing the sum of the search and action execution.

    The comparative analysis done in \cite{arica2017empirical} considers only a static environment, neglecting the possible variations in the algorithm performance under different environmental characteristics. This limitation hampers us from creating a clear understanding of different search algorithms that could perform under different environmental characteristics, precisely in the context of path planning. Similarly, the study conducted in \cite{koenig2009comparing} restricts its investigation to only two algorithms each one from a different class i.e., real-time and incremental heuristic search. Even though it provides an insightful comparison, the study does not fully capture the breadth of available algorithms in these classes.
    
    Moreover, the study aimed to provide a recommendation on when to use each of the algorithm classes, but also does not investigate algorithm performances in diverse environmental characteristics.

    Consequently, considering the challenge posed by the presence of a wide range of search algorithms and the lack of a comprehensive comparative analysis under different environment settings, we propose to evaluate and analyze the performance of some search algorithms under different environment characteristics in the context of path planning, But first, we define the characteristics of the environment used, and through our experiments, we aim to provide a comprehensive evaluation of these algorithms followed by proposing our selection algorithm.

\section{Experimental environment and performance metrices}
    Heuristic search algorithms play an important role in fields such as robotic pathfinding, as they determine the optimal path given a starting position and a goal position. Grid-based environments are commonly used for representing real-world environment scenarios, where these algorithms can be implemented, such as in autonomous navigation and robotics \cite{30720}.
    In this study, we comprehensively analyze well-known heuristic search algorithms, namely D*, D* Lite, LPA*, LRTA*, RTAA*, and ARA* in different grid environments. We use the Euclidean distance heuristic to guide the search of the algorithms and assess the impact of a few grid characteristics, such as the obstacle density and the grid size, on the performance of the algorithms.    

    In our study, we use grid-based environments due to their simplicity, and control ease, in addition to being commonly used in path-planning tasks in the research community. The grids are composed of white cells, representing traversable states, whereas the black ones represent non-traversable obstacles. The agent in our simulation can move in eight directions, with a cost equal to 1 for horizontal and vertical moves, and $\sqrt{2}$ for diagonal movements. We used two types of grid environments: randomly generated grid environments and personalized grid environments. 
        
        \paragraph{Randomly generated grid-based environments:}
        
        The grids are characterized by three parameters: grid size $(NxN)$, obstacle density, and the distance between the start and the goal states. To investigate the impact of each parameter on the performance of the search algorithms, we varied each parameter independently while keeping the other parameters constant. The variations included varying the grid size, varying start to goal distance, and varying the obstacle densities. For each grid in a variation, we generated ten random instances of the same grid parameters (e.g., a grid with 0.25 of obstacle density, size 300x300, and 140 of start to goal distance, have ten instances, which were generated randomly). 

        \paragraph{Personalised grid environment:}
        Designed for simulating more specific scenarios. These grids have fixed size (71x31 units) and a fixed position for both the start and goal states, and they were divided into two parts based on their obstacle configuration:
        \begin{itemize}
        \item \textbf{Horizontal wall configuration}: For these experiments, we add horizontal walls of half grid width every 10 units of grid length. We added the walls in two orientations: once from left to right, and once from right to left in the newly generated grid.
        \item \textbf{Horizontal wall length configuration}: Here, we add all possible walls that can be placed within the grid length, and each time we generate a new grid we increase all wall lengths by 2 units.
        \end{itemize} 

     We adopted these two distinct environments to provide a thorough analysis, seeking to reveal nuanced insights into the performance of the search algorithm in the presence of two different hindering scenarios. In the first one, obstacles are scattered randomly within the grid, whereas in the second one, the wall-like structures, appear as a mass of connected obstacles.

    To evaluate the performance of the different search algorithms used in the experiments, we have selected the following metrics:
    \begin{itemize}
        \item \textbf{Path cost:} The metric measures the path length or the number of executed actions from the start to the goal state. It indicates the quality of the solution.
        \item \textbf{Memory consumption:} It measures the required amount of memory for the algorithm to find a solution. It is relevant to check the scalability of the algorithm, and it is measured in (KB). 
        \item \textbf{Solving Time:} Represent the total time an algorithm takes to find a solution in (ms), measured in milliseconds (ms).
    \end{itemize}
    We carried out our experiments on 3.30GHz 27 Intel i9 cores, equipped with 250Gb of RAM and running Ubuntu Linux 22.04. Using the following settings: 
    All algorithms were using the Euclidean distance as the heuristic function. For LRTA* and RTAA*, we set the number of expended nodes to 250  for each iteration. The ARA* algorithm was run with a weight of 2.5 for the heuristic.
    We ran each algorithm 100 times on each grid to account for randomness and to ensure the reliability of the results. 
        
        \section{Experimental results}
\begin{figure}[H]
                \centering
                \includegraphics[width=1.15\linewidth]{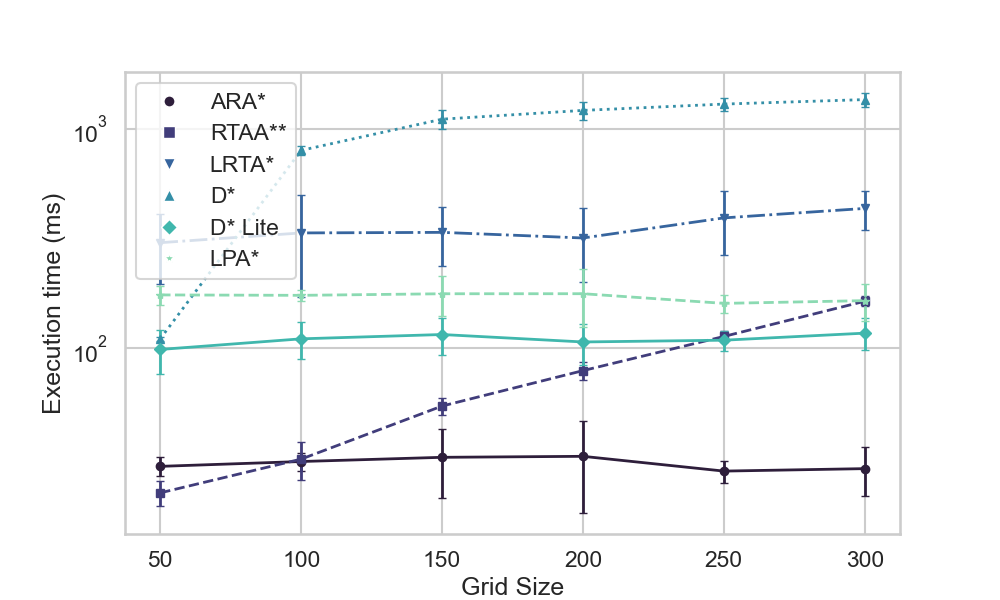}
                \caption{solving time vs. Grid Size}
                \label{fig:executionGrid}
            \end{figure}

\begin{figure}[H]
                \centering
                \includegraphics[width=1.15\linewidth]{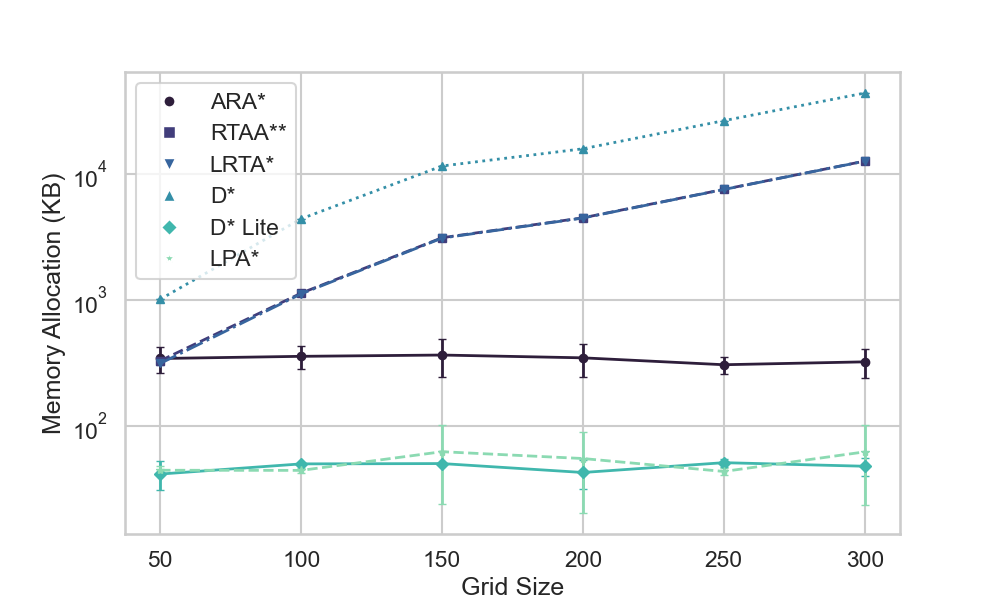}
                \caption{Memory Allocation vs. Grid Size}
                \label{fig:memoryGrid}
            \end{figure}

\begin{figure}[H]
            \centering
            \includegraphics[width=1.15\linewidth]{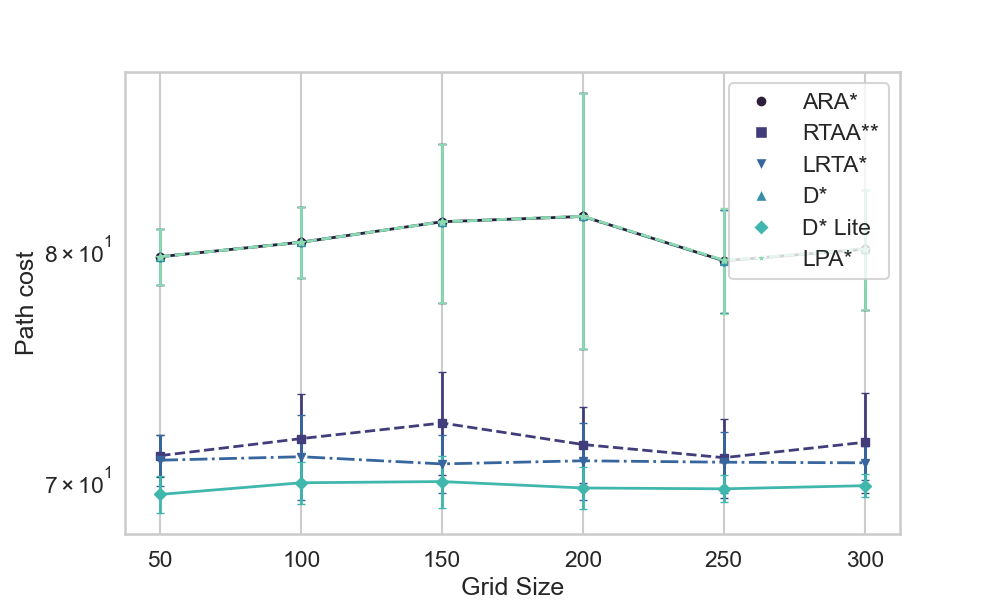}
            \caption{Path Cost vs. Grid Size}
            \label{fig:pathcostGrid}
        \end{figure}

            \paragraph{Grid Size Variation:} 
            In the results obtained by varying the grid size (see Figure \ref{fig:executionGrid}, \ref{fig:pathcostGrid}, \ref{fig:memoryGrid}), ARA* displays relatively stable solving time. However, it has some fluctuation in its standard deviation, with a slight increase as the grid size increases. Its maximum value of (76ms) was obtained at size 200. At grid size 50, RTAA* was the fastest algorithm, recording a solving time of (21ms). Subsequently, its solving time and allocated memory increased notably with the grid size. which indicates that the grid size affects the performance of RTAA* as it increases; and the algorithm is forced to do more searches.
            LRTA* has the highest solving time across all sizes compared to ARA*, RTAA*, D* Lite, LPA*, and even D* at size 50. Its performance greatly varies as indicated by its high standard deviations. Moreover, LRTA* did not show a clear trend in increasing time relative to grid size, indicating that it may be unpredictable and inefficient for this task.
            Both D* Lite and LPA* exhibited relatively stable performances. D* Lite has a lower solution time across all grid sizes compared to LPA*. However, both algorithm's standard deviations show some degree of fluctuation. D*, on the other hand, showcased an extreme increase in solution time when transitioning from size 50 to 100, revealing its challenges as the grid size increases. 
            
            Regarding path cost (see Figure \ref{fig:pathcostGrid}), while LPA*, D*, and ARA* lines overlap, suggesting similar path costs across all grid sizes,  D* Lite consistently generates the shortest paths across all grid sizes, with a relatively small standard deviation. In terms of memory allocation (see Figure \ref{fig:memoryGrid}), LPA* and D* Lite were consistent across all grid sizes, requiring the least memory among all algorithms. In contrast, RTAA*, LRTA*, and D* demanded more memory as the grid size increased.
            \begin{multicols}{3}

            \end{multicols}

\begin{figure}[H]
            \centering
            \includegraphics[width=1.13\linewidth]{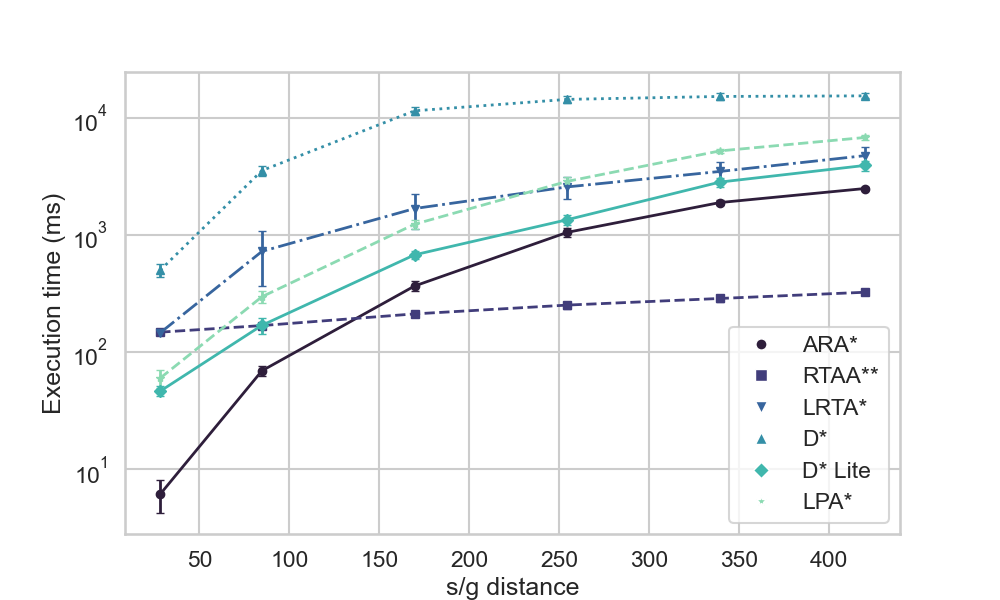}
            \caption{solving time vs. SG Distance}
            \label{fig:executionSg}
        \end{figure}

\begin{figure}[H]
    \centering
    \includegraphics[width=1.13\linewidth]{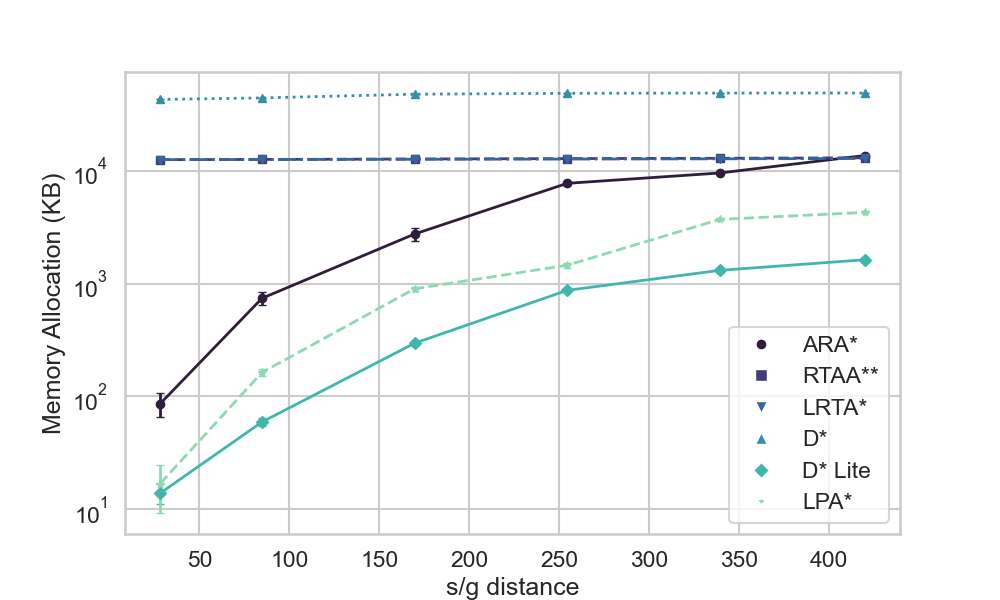}
    \caption{Memory Allocation vs. SG Distance}
    \label{fig:memorySg}
\end{figure}

\begin{figure}[H]
    \centering
    \includegraphics[width=1.13\linewidth]{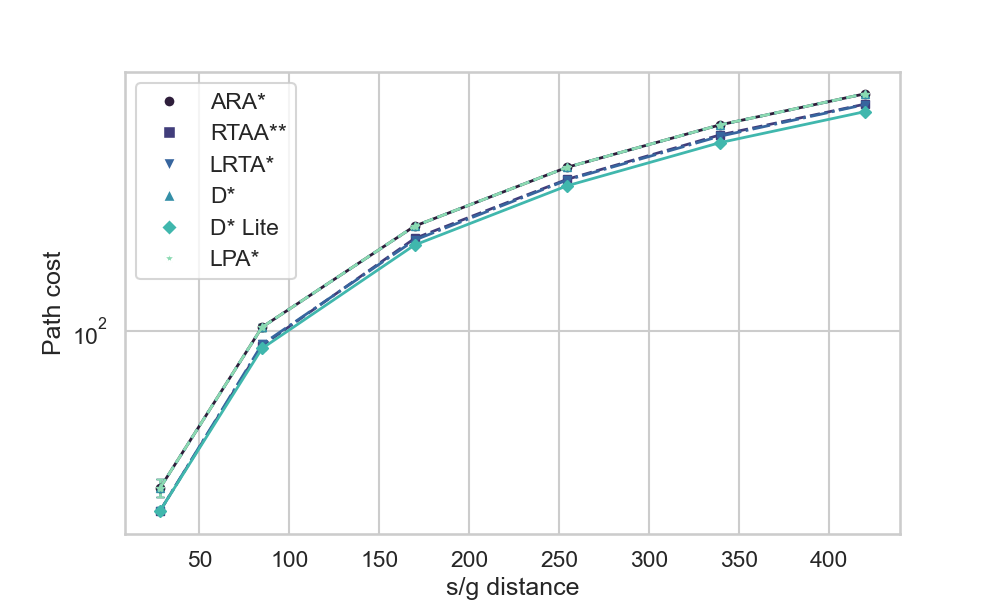}
    \caption{Path Cost vs. SG Distance}
    \label{fig:pathcostSg}
\end{figure}

            \paragraph{Start to goal distance variation:}
            The obtained results from running the algorithms on grids with varying start-to-goal distances (see figures \ref{fig:executionSg}, \ref{fig:memorySg}, \ref{fig:pathcostSg}) revealed that both the mean of the path cost and the mean of solving time for all algorithms escalate as the distance increases (see Figure \ref{fig:executionSg}, \ref{fig:pathcostSg}). This outcome was expected, since longer distances naturally demand more computational efforts and more nodes to expand.

            ARA* seems to have a strong correlation with the distance between the start and goal; its solving time is drastically influenced by this factor. Precisely, ARA* remains the fastest algorithm for distances smaller than approximately 140. Beyond this threshold, however, RTAA* takes the lead in terms of solving time. 
            
            Observing the path cost (see Figure \ref{fig:pathcostSg}), the lines for D* Lite, LRTA* RTAA* overlap, indicating similar performances. Correspondingly, LPA*, ARA*, and D* also exhibit overlapping lines, where D* Lite being the algorithm with the lowest path cost. In the context of allocated memory (see Figure \ref{fig:memorySg}), D* lite allocates the least memory for all distances followed by LPA* and then ARA*. The rest of the algorithms allocate almost a similar amount of memory with D* being the worst. 
                     \begin{multicols}{3}
        \end{multicols}

\begin{figure}[H]
                    \centering
                    \includegraphics[width=1.15\linewidth]{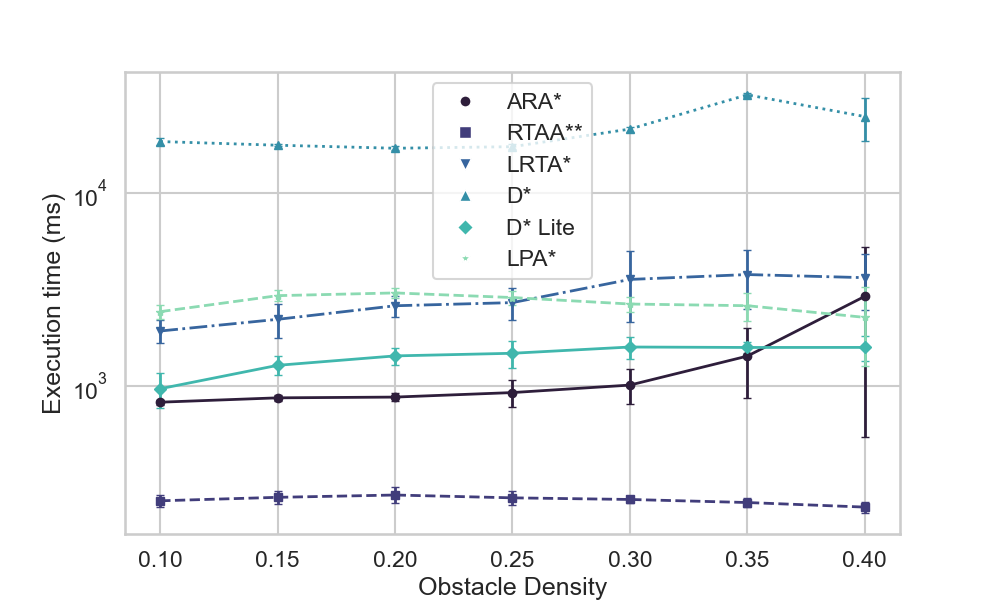}
                    \caption{solving time vs. Obstacle Density}
                    \label{fig:executionObstacle}
                \end{figure}

            \paragraph{Obstacle density variation:}
            RTAA* constantly excels by producing the shortest solving time among all algorithms and across all obstacle densities as depicted in Figure \ref{fig:executionObstacle}. This superior performance can be explained due to its rapid heuristic values update procedure within its local search space. The rest of the algorithm's solving time increases as the obstacle density rises, with D* being the one with a higher solving time, except for LPA*, which starts to decrease slightly after a density of 0.20. 

\begin{figure}[H]
                    \centering
                    \includegraphics[width=1.15\linewidth]{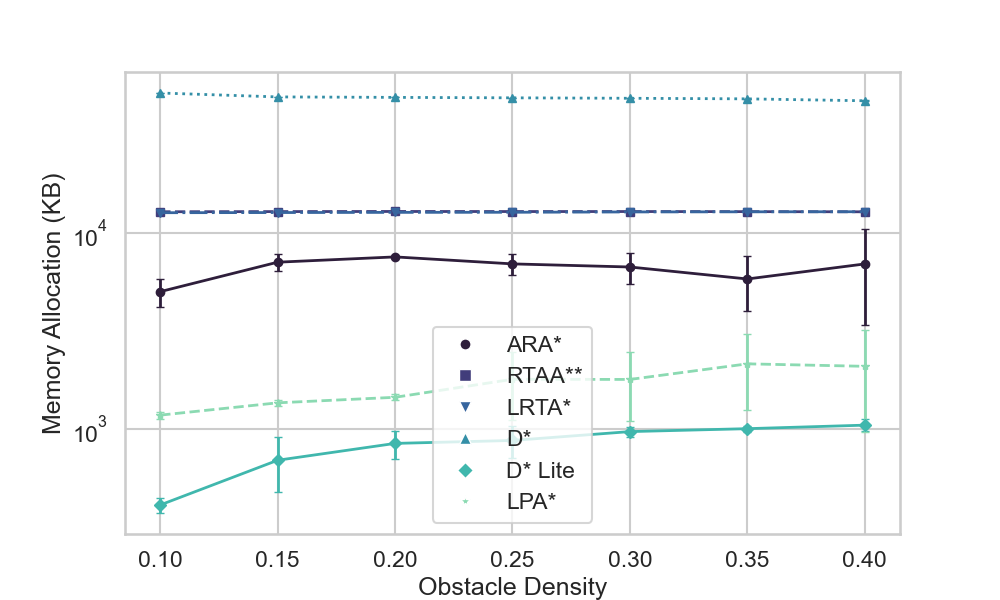}
                    \caption{Memory Allocation vs. Obstacle Density}
                    \label{fig:memoryObstacle}
                \end{figure}

\begin{figure}[H]
                    \centering
                    \includegraphics[width=1.15\linewidth]{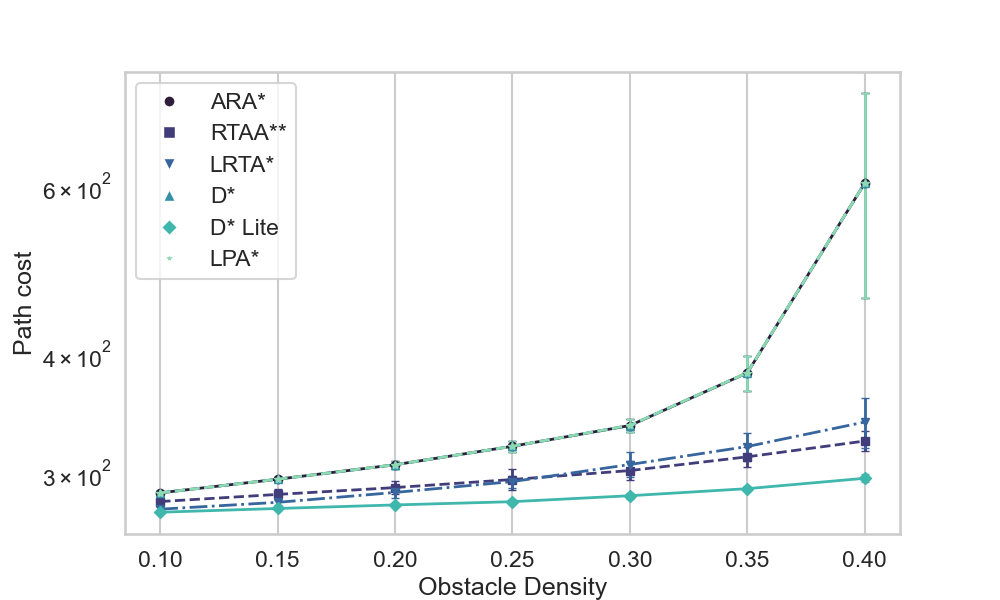}
                    \caption{Path Cost vs. Obstacle Density}
                    \label{fig:pathcostObstacle}
                \end{figure}

            Figure \ref{fig:pathcostObstacle} shows that the path cost of all algorithms tends to increase as the density increases, which can be a predictable outcome since denser environments pose more complex navigation challenges. 
            Amongst all algorithms, D* Lite maintains the lowest path across all densities, marking its efficiency in complex environments. D* Lite's performance is followed by LRTA* for obstacle densities below 0.25, and RTAA* outperforms the rest for densities higher than 0.25. In addition to maintaining the lowest path cost, D* Lite allocates the least amount of memory at all densities, as depicted in Figure \ref{fig:memoryObstacle}, followed by LPA*. In contrast, both RTAA* and LRTA* consume a lot of memory, but not as much as D*, which allocates even more.  

            \begin{multicols}{3}
                
        
            \end{multicols}
            We hypothesize that changing other parameters that we have fixed while varying other ones could indeed provide us with further insight into how these parameters interact and affect the performance of the search algorithms i.e., using an obstacle density of 0.4 rather than 0.25 while changing the grid size. 
            However, to maintain simplicity and manage the computational resources, we opted to keep a balanced grid size, obstacle density, and distance from the start to the goal that will help us fairly represent an environment for a pathfinding task. 
            Also, the combination of varying one parameter independently from the other ones and then repeating the same experiment, in which we vary the fixed parameter using all other possible values would dramatically increase the number of possible experiments that must be done, in addition to the number of runs that they must be made for each grid, which will amplify the number of experiments that must be performed.

\begin{figure}[H]
                    \centering
                    \includegraphics[width=1.15\linewidth]{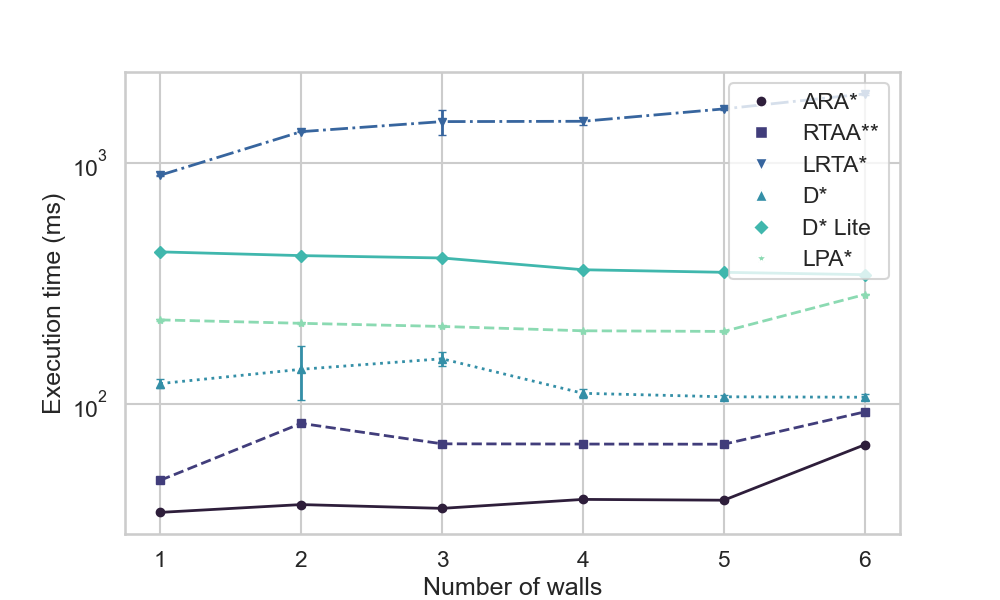}
                    \caption{solving time vs. Number of walls}
                    \label{fig:executionNumberOfWalls}
                \end{figure}

\begin{figure}[H]
                    \centering
                    \includegraphics[width=1.15\linewidth]{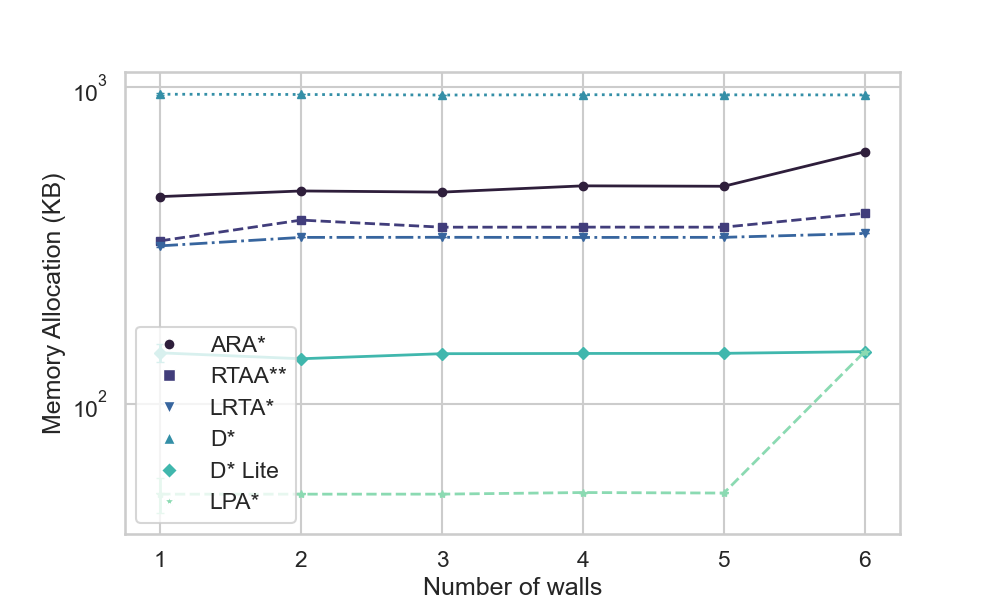}
                    \caption{Memory Allocation vs. Number of walls}
                    \label{fig:memoryNumberOfWalls}
                \end{figure}

\begin{figure}[H]
                    \centering
                    \includegraphics[width=1.15\linewidth]{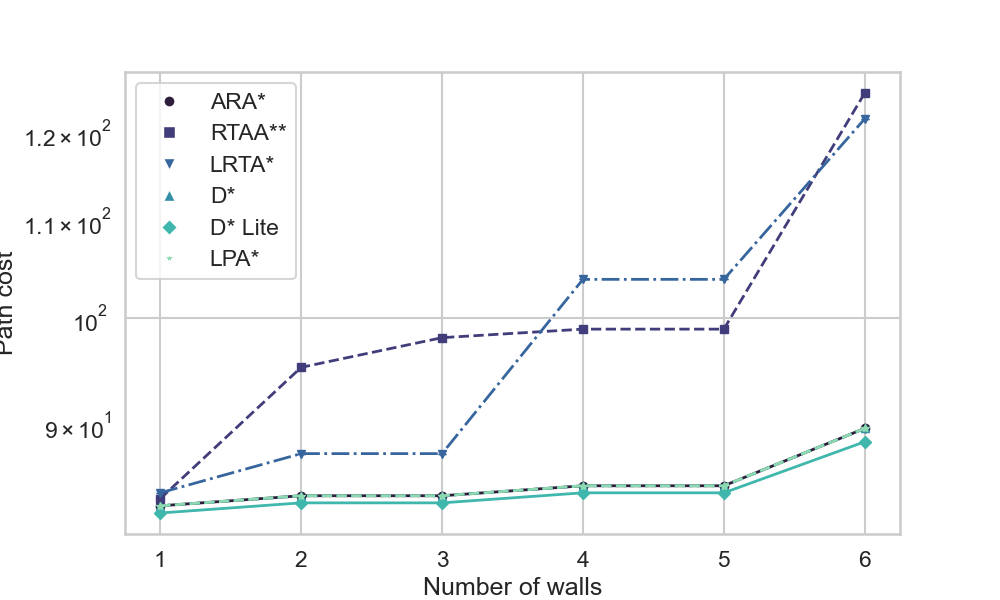}
                    \caption{Path Cost vs. Number of walls}
                    \label{fig:pathcostNumberOfWalls}
                \end{figure}

            \paragraph{Horizontal wall configuration:}
            In the horizontal wall configuration results, as depicted in Figures (\ref{fig:executionNumberOfWalls}, \ref{fig:memoryNumberOfWalls}, and \ref{fig:pathcostNumberOfWalls}). We observed that the path cost tends to increase for all algorithms as walls are added (see Figure \ref{fig:pathcostNumberOfWalls}), which is expected. 
            LRTA* exhibited the highest solving time (see Figure \ref{fig:executionNumberOfWalls}), followed, in descending order by D* Lite, RTAA*, LPA*, ARA*, and LRTA*. 
            Regarding the obtained results for the path cost, D* Lite produces the most optimal paths followed by ARA*, D*, and LPA*. 
            In terms of memory allocation (see Figure \ref{fig:memoryNumberOfWalls}), all algorithms tend to allocate the same amount of memory even when new walls are introduced, with only slight increases. Among all algorithms, D* allocates the highest amount of memory, while LPA* allocates the least memory, followed by D* Lite, LRTA*, RTAA*, and ARA*. However, it is worth noting that both ARA* and LPA* had a remarkable increase in memory consumption when adding the last wall. 
            \begin{multicols}{3}
                
\begin{figure}[H]
                    \centering
                    \includegraphics[width=1.15\linewidth]{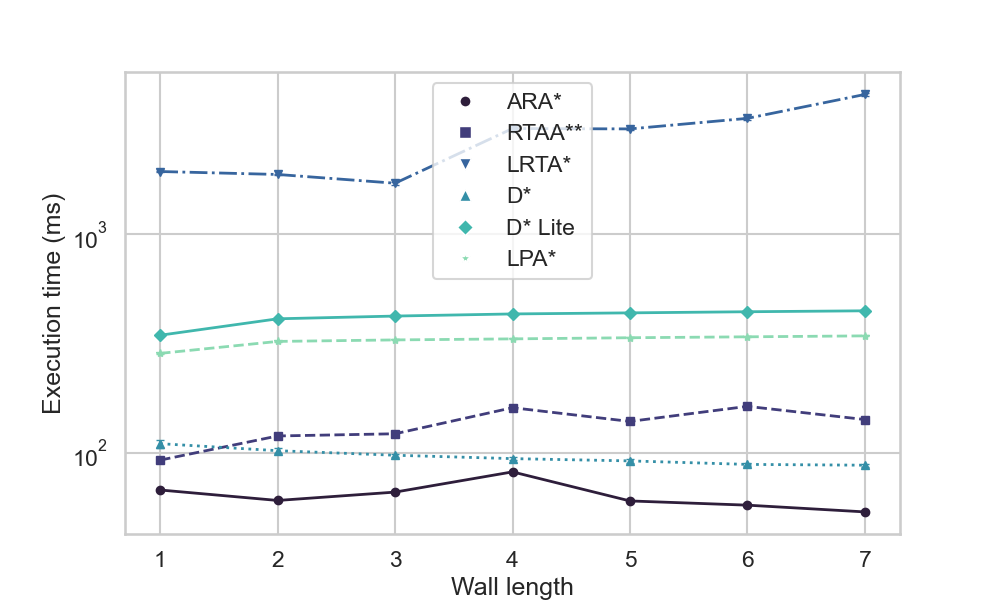}
                    \caption{solving time vs. Walls length}
                    \label{fig:executionWallsLength}
                \end{figure}

\begin{figure}[H]
                    \centering
                    \includegraphics[width=1.15\linewidth]{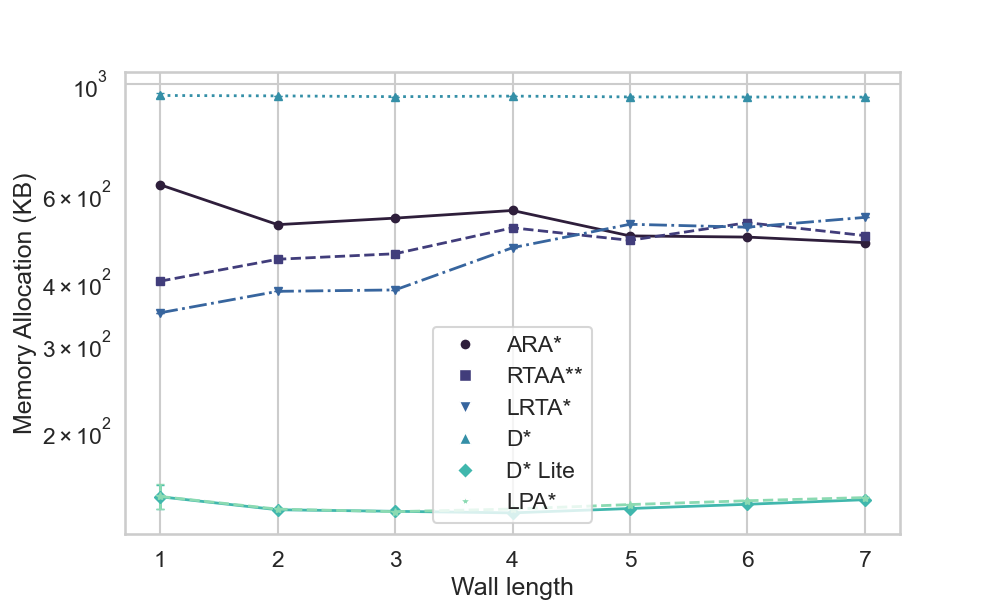}
                    \caption{Memory Allocation vs. Walls length}
                    \label{fig:memoryWallsLength}
                \end{figure}

\begin{figure}[H]
                    \centering
                    \includegraphics[width=1.15\linewidth]{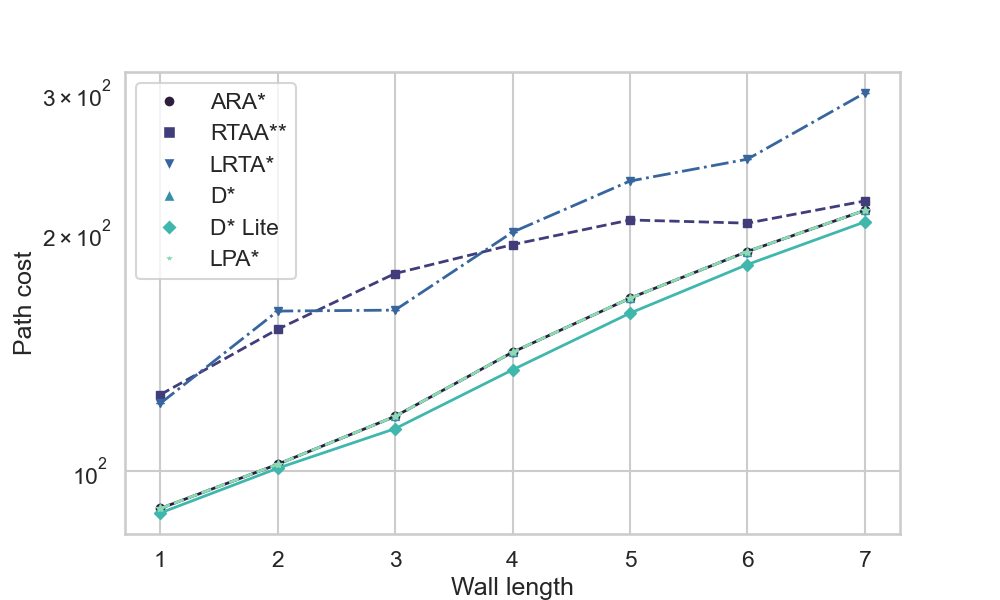}
                    \caption{Path Cost vs. Walls length}
                    \label{fig:pathcostWallsLength}
                \end{figure}

            \end{multicols}
            \paragraph{Horizontal wall length configuration:}
            The results obtained from varying the wall sizes are depicted in the figures \ref{fig:executionWallsLength}, \ref{fig:memoryWallsLength}, and \ref{fig:pathcostWallsLength}. The numbers on the axis represent seven different lengths. The initial length corresponds to half the grid width, and with each subsequent step, it increases by two obstacles.

             All algorithm's solving times varied across the different wall lengths showing a general trend of increasing as the wall lengths increase as depicted in Figure \ref{fig:executionWallsLength}. LRTA* recorded the highest solving times, whereas ARA* recorded the lowest across all wall lengths. For the amount of memory used by the algorithms (see Figure \ref{fig:memoryWallsLength}), D* was again the one that consumed more memory. In contrast, both D* Lite and LPA* allocated almost similar and the latest amount of memory over all seven wall lengths. Turning to the path cost metrics (see Figure \ref{fig:pathcostWallsLength}), all algorithm's path costs tend to increase as the wall lengths increase, with D* Lite generating the lowest path costs for all wall lengths. 

            \begin{multicols}{3}
        
            \end{multicols}

            Based on the results obtained in the grid size variation, the performance of most algorithms, particularly RTAA*, was affected by grid size. In the meanwhile, the high level of consistency of ARA* performance in terms of solving time regardless of the grid size indicates its suitability for various grid sizes higher than 100. D* proved to be the one generating the lowest path costs, and also allocating the least memory alongside LPA*.
            Obstacle density results showed that it is a factor that influences the performance of all algorithms. However, D* Lite kept generating the most efficient paths, which indicates its effectiveness in dense environments.
            Based on the results obtained by increasing the start-to-goal distance, ARA* appears to be the most affected one since its solving time continuously increases. In addition to allocating the least memory, D* Lite consistently exhibits the optimal paths, which indicates its utility where the shortest path is of importance.
            Adding horizontal walls each time in the same grid setting has increased the path length and solving time for all algorithms with D* being the worst. D* Lite keeps its optimal performance by generating the most optimal paths, which suggests its suitability in environments with many obstacles. However when increasing the length of the walls, ARA* displayed the lowest solving times, suggesting its efficiency in environments with extensive barriers. Moreover, D* Lite keeps generating the lowest path costs.
            
            In summary, while each algorithm has its strengths and weaknesses, D* Lite has continuously shown a good performance across most conditions, particularly in generating the optimal paths. Meanwhile, ARA* has proved its stability in producing the fastest paths regardless of the grid size. Moreover, RTAA* showed its ability to generate faster paths regardless of the obstacle density due to its faster procedure in updating the heuristic values.

    \section{Selection algorithm and example of execution}

    Based on the insights derived from our experimental results, we propose the selection algorithm represented in Algorithm \ref{alg:algorithmselection}. 
    The algorithm is designed to select the appropriate search algorithm based on various priorities alongside the characterization of the environment used.
    
    The selection algorithm emphasizes the desired priority first, which could manifest in various aspects of pathfinding tasks, including the path cost, memory usage, or the time taken to find a solution. 
    The rational reason for emphasizing $"$Priority$"$ at the beginning of the selection algorithm is that, based on this user choice, different search algorithms may be considered to suit best the addressed problem. Thus, by tackling the $"$Priority$"$  upfront, The selected algorithm will eventually, cater to the main requirements of the task at hand. 

    The algorithm takes as input the Grid that is of size $NxN$ with the start and goal positions, the distance threshold, and the priority criterion. If we aim to minimize memory usage, path cost, or both (Line 2), the algorithm suggests using D* Lite. However, If minimizing the solving time is our primary concern (line 4), we should first calculate the Euclidean distance between the start and the goal using the function in line 12. If the distance is higher or equal to the threshold (in our experiments it is equal to 140), the algorithms suggest using RTAA*. If the distance is less, then the selection algorithm suggests using ARA*.

\begin{algorithm}
\caption{selection algorithm}
\label{alg:algorithmselection}
\begin{algorithmic}[1]
\newcommand{\Input}{\item[\textbf{Input:}]} 
\newcommand{\Output}{\item[\textbf{Output:}]} 

\Input 
    \Statex Grid: < array:  \( size[N][N] \), 
    start\( (start_x, start_y) \),
    goal\( (goal_x, goal_y) \)>
    \Statex Integer: \( D \)      \Comment{Distance threshold} 
    \Statex String: P          \Comment{Priority_criterion}

\Output 
    \Statex Selected algorithm based on specified criteria.

\Function{select_algorithm}{ Grid<size, start, goal>, D, P}
    \If{ P is `Memory' \textbf{or} P is `PathCost'}
        \State \Return \textproc{dstar_lite_algo}
    \ElsIf{P is `SolvingTime'}
        \If{\Call{ComputeEuclideanDistance}{start, goal} \(\geq D\)}
            \State \Return \textproc{rtaa_algo}
        \Else
            \State \Return \textproc{ara_star}
        \EndIf
    \EndIf
\EndFunction
\Function{ComputeEuclideanDistance}{start, goal}
    \State \Return \(\sqrt{(start_x - goal_x)^2 + (start_y - goal_y)^2}\)
\EndFunction
\end{algorithmic}
\end{algorithm}

\begin{table}
\centering
\caption{Selection algorithm evaluation results}
\label{tab:combinedComparison}
\adjustbox{max width=\textwidth}{
\begin{tabular}{|c|c|c|c|c|c|c|c|c|}
\hline
\multicolumn{9}{|c|}{ Randomly generated NxN grid environments} \\
\hline
Algorithm & Number of walls & Wall length & Obstacle density & Grid size & s/g distance & Path cost & Memory Allocation (KB) & Solving time  (ms) \\
\hline
RTAA* & - & - & 0.35 & 100 & 43.174 & 81.012 & 1145.648 & 86.106 \\
ARA* & - & - & 0.35 & 100 & 43.174 & 86.041 & 437.012 & \cellcolor{blue!25}63.376 \\
D* Lite & - & - & 0.35 & 100 & 43.174 & \cellcolor{yellow!25}73.698 & \cellcolor{green!25}109.248 & 252.110 \\
\hline
RTAA* & - & - & 0.35 & 250 & 56.080 & 78.769 & 7520.803 & 165.108 \\
ARA* & - & - & 0.35 & 250 & 56.080 & 83.798 & 173.603 & \cellcolor{blue!25}33.001 \\
D* Lite & - & - & 0.35 & 250 & 56.080 & \cellcolor{yellow!25}73.698 & \cellcolor{green!25}94.861 & 196.605 \\
\hline
RTAA* & - & - & 0.25 & 100 & 38.118 & 77.012 & 1159.855 & 111.278 \\
ARA* & - & - & 0.25 & 100 & 38.118 & 75.455 & 382.359 & \cellcolor{blue!25}55.283 \\
D* Lite & - & - & 0.25 & 100 & 38.118 & \cellcolor{yellow!25}72.870 & \cellcolor{green!25}111.344 & 285.450 \\
\hline
RTAA* & - & - & 0.35 & 100 & 36.069 & 70.497 & 1119.316 & 55.832 \\
ARA* & - & - & 0.35 & 100 & 36.069 & 97.426 & 825.953 & \cellcolor{red!25}136.500 \\
D* Lite & - & - & 0.35 & 100 & 36.069 & \cellcolor{yellow!25}69.083 & \cellcolor{green!25}91.369 & 218.514 \\
\hline
\multicolumn{9}{|c|}{Personalised grid environments} \\
\hline
Algorithm & Number of walls & Wall length & Obstacle density & Grid size & s/g distance & Path cost & Memory Allocation (KB) & Solving time  (ms) \\
\hline
RTAA* & 6 & 25 & - & 31x71 & 73.539 & 177.480 & 368.113 & 131.508 \\
ARA* & 6 & 25 & - & 31x71 & 73.539 & 117.195 & 531.484 & \cellcolor{blue!25}88.499 \\
D* Lite & 6 & 25 & - & 31x71 & 73.539 & \cellcolor{yellow!25}113.095 & \cellcolor{green!25}243.723 & 708.3793 \\
\hline
RTAA* & 6 & 16 & - & 31x71 & 73.539 & 83.698 & 334.370 & 64.666 \\
ARA* & 6 & 16 & - & 31x71 & 73.539 & 83.113 & 449.302 & \cellcolor{blue!25}46.540 \\
D* Lite & 6 & 16 & - & 31x71 & 73.539 & \cellcolor{yellow!25}82.527 & \cellcolor{green!25}244.100 & 686.459 \\
\hline
\end{tabular}
}
\end{table}

\subsection{Example of execution}

To showcase the efficiency of our selection algorithm, we have designed an execution example that consists of generating a grid with random parameters i.e., choose a random grid size, obstacle density, and start to goal distance. After that, we generate an identical grid, however, we change only one parameter each time while keeping the other parameters as they were in the initial grid. We do the same for generating personalized grid environments, where we generate a grid with a random choice of wall number, and then we change the wall lengths.
 
We run RTAA*, ARA*, and D*Lite algorithms on the generated grids (all obtained results are shown in table \ref{tab:combinedComparison}) alongside our selection algorithm, so that we can compare the obtained results with what the selection algorithm is suggesting to use for the given grid.

 It is essential to highlight that the randomly chosen values for the grid parameters are all within the range of the values used in the experiments, which ensures that the thresholds chosen are relevant. Also, the reason behind using only RTAA*, ARA*, and D*Lite in this execution example is because of their standout performance in our experiments.

For each grid, we run our selection algorithm each time with a different priority, and it suggests using an algorithm that will perform the best given the selected priority. The highlighted values in the table refer to the outcomes derived from the suggested search algorithm by our selection algorithm. These highlighted values are notably the best that we can obtain for each grid given a certain priority except for the red one; yellow represents the most efficient paths, green represents the best values that we can obtain if we want to reduce memory consumption, and the values highlighted in purple indicates the best solving time.
However, our algorithm failed in selecting the right algorithm to use in the fourth grid when prioritizing the solving time, where the shortest solving time was found by RTAA*, instead our algorithm suggested using ARA*.

\section{Conclusion}

    This work aimed to evaluate the performance of several known heuristic search algorithms, such as D*, D* Lite, LPA*, LRTA*, RTAA*, and ARA* in terms of solving time, memory consumption, and path cost. 
    The evaluation was made using different randomly generated grid environments with different characteristics alongside personalized grid environments with horizontal walls and different wall lengths as different hinders.

    Our experimental evaluation revealed that all the algorithms exhibit different performances with strengths and weaknesses under different grid characterizations. D* Lite consistently generated the shortest paths even in obstacle-dense grids, indicating its efficiency in dense environments. ARA* consistently provides faster solutions as the grid size increases, particularly larger than 100, while RTTA* generates faster solutions in smaller grid sizes, with the advantage of not being affected by dense environments.

    Our study provides valuable insights into selecting the appropriate heuristic search algorithm in the pathfinding domain. Using these insights, we propose a selection algorithm used to optimize the performance needed in a pathfinding domain, such as, solving time, path length, or memory consumption.
    However, our evaluation focused only on static environments, while dynamic environments may introduce additional challenges. Furthermore, we considered a limited set of experiments, not covering all possible combinations of the grid characteristics. This limitation means that the selection algorithm might not always recommend the most optimal solution, as seen in the example we introduced to evaluate our algorithm.  
        
        In future work, we aim to address these limitations as follows:
    Firstly, we plan to extend our evaluation of the search algorithms to include dynamic environments. 
    Secondly, we intend to explore various combinations of both domain characterization and priorities.
    Furthermore, we want to include additional scenario configuration, extending the random obstacles and the walls scenarios. With such additional understanding, we aim to refine our selection algorithm to automatically take decisions among the best search algorithms, based on the type of obstacles in the local search space.


\bibliographystyle{eptcs}
\bibliography{generic}


\end{document}